\documentclass{webofc}

\usepackage[varg]{txfonts}
\usepackage{hyperref}
\usepackage{url}
\usepackage{orcidlink}
\usepackage{listings}
\usepackage{booktabs}
\usepackage{multicol}
\hypersetup{colorlinks=true,citecolor=blue,urlcolor=blue,linkcolor=blue}

\begin{document}
\title{Hypothesis-awkward}
\subtitle{Property-Based Testing Strategies for Awkward Array}

\author{\firstname{Tai} \lastname{Sakuma}\inst{1}\orcidlink{0000-0003-3225-9861}\fnsep\thanks{\email{sakuma@princeton.edu}} \and
        \firstname{Ianna} \lastname{Osborne}\inst{1}\orcidlink{0000-0002-6955-1033} \and
        \firstname{Peter} \lastname{Elmer}\inst{1}\orcidlink{0000-0001-6830-3356}
}

\institute{Princeton University, Princeton, NJ, USA}

\abstract{%
\emph{Hypothesis-awkward} is a collection of Hypothesis strategies for Awkward
Array. Awkward Array can represent a wide variety of nested, variable-length,
mixed-type data. Many tools that process Awkward Arrays are widely used and
actively developed. The unit test cases of many of these tools list predefined
input samples. In practice, such samples can cover only a small portion of the
vast combinatorial space of Awkward Array instances. Hypothesis, a Python
property-based testing library, generates test data that can make test cases
fail and automatically explores edge cases. The main strategy of
Hypothesis-awkward generates nearly all possible Awkward Arrays, with options to
control the layout, data types, missing values, masks, and other array
attributes. Property-based tests written with these strategies have begun to run
in the continuous integration of Awkward Array; to date, they cover a small part
of Awkward Array and have found 37 bugs in it, as well as one in Hypothesis
itself. These strategies make the tests of Awkward Array and of the tools that
use it more reliable. }
\maketitle
\section{Introduction}
\label{sec-intro}

\subsection{Reliable tests and iterative development}
\label{sec-reliable}

The requirements of software change over time: for example, additional features
requested by users, support for different hardware, and interfaces for other
kinds of extensions. Software that remains in use has to be changed continually
to keep satisfying its requirements.

In iterative development, software is changed in cycles of refactoring and
feature changes. Refactoring, without changing the expected behavior of the
code, makes the next feature change easier to implement~\cite{fowler2018}.
Today, AI coding agents can shorten the cycle. After each refactoring and each
feature change, all tests must pass.

This cycle depends on reliable tests. They guarantee that a refactoring leaves
the expected behavior unchanged and that a feature change alters the behavior
only where the expectation has changed; then, the change can be released.
Reliable tests include unit tests that cover nearly all of the code and any
necessary integration, system, and regression tests. They must explore the edge
cases, covering a set of cases broader than the set of actual use cases.
Exploring the edge cases is the subject of this paper.

\subsection{Example-based tests}
\label{sec-example-based}

Example-based tests assert specific outputs for specific inputs. Many software
packages rely on such unit tests, with input samples that are predefined: chosen
by developers or generated by AI, and hard-coded into the test cases.

Predefined samples are sufficient when the edge cases of the input can be
enumerated. When the input is a complex data structure, enumerating the edge
cases becomes impractical. As a result, the tests might not find bugs in edge
cases: they pass on the predefined samples while the code does not behave as
expected on other inputs.

Predefined samples also allow an implementation to be fitted to them. In
test-driven development, the implementation is written to make the tests pass,
and its author can read the samples. For example, an AI coding agent can
generate code that returns the asserted outputs for the predefined samples
without implementing the required behavior.

Whether edge cases are missed or the implementation is fitted to the samples,
the tests do not guarantee that the code behaves as expected; they are not
reliable.

\section{Complex data structures in Awkward Array}
\label{sec-awkward}

Awkward Array~\cite{awkward}, a Python library for arrays, can represent complex
data structures: nested lists, lists of variable length (jagged arrays), missing
values, records, and values of mixed types.

An array is described at three levels: type, form, and layout. The \emph{type}
describes the structure of the data. Listing~\ref{lst-types} shows examples of
the types; the numeric types, such as \texttt{int64}, are NumPy~\cite{numpy}
data types (dtypes). The types can be nested in one another to any depth. The
\emph{form} is a node of a tree that describes how the data are stored. The
\emph{layout} is a node of a tree of the same structure that holds the data;
each node is an instance of one of the twelve subclasses of
\texttt{ak.contents.Content}~\cite{awkward}. Arrays with the same type and
values can have different layouts: missing values can be represented by any of
four subclasses, the offsets of a list need not start at zero, a layout can hold
data that no element refers to, and the data can be virtual, that is, not yet
materialized.

\begin{lstlisting}[caption={Examples of Awkward Array data types.},
  label={lst-types}, xleftmargin=0.2\linewidth, xrightmargin=0.2\linewidth]
5 * int64                      # One-dimensional
3 * 5 * float64                # Rectilinear
3 * var * int32                # Jagged
3 * ?float64                   # Optional (nullable)
3 * var * ?int64               # Jagged optional
3 * {x: float64, y: int32}     # Records
3 * union[int64, string]       # Heterogeneous
3 * var * {x: float64,
           y: var * int32}     # Nested combination
\end{lstlisting}

The possible arrays constitute a vast combinatorial space. Each node of a layout
is one of twelve kinds, each leaf has one of many dtypes, and each list has any
length, including zero. These choices multiply with every level of nesting. A
tool is expected to work on every array of the kinds that it accepts. Predefined
test samples can in practice cover only a small portion of this space.

\section{Property-based tests and Hypothesis}
\label{sec-pbt}

Property-based tests use generated samples instead of predefined ones. The
expected output for a generated sample is generally not known, so the test
cannot assert it. Instead, it asserts that the code satisfies expected
\emph{properties} for all inputs that meet certain conditions. For example, a
function that sorts a list of mutually comparable elements must reorder them so
that each is no greater than the next.

\emph{Hypothesis}~\cite{MacIver2019Hypothesis} is a property-based testing
library for Python. A \emph{strategy} generates random samples that meet given
conditions in search of one that makes a test fail. A test can \emph{draw} many
different samples from a strategy: hundreds, thousands, or more. When a test
fails, Hypothesis \emph{shrinks} the failing sample; that is, it searches for
the simplest sample that causes the same failure.

Hypothesis automatically explores the edge cases that reliable tests must
explore (Sect.~\ref{sec-reliable}). The samples are generated at run time, so an
implementation cannot be fitted to them. Property-based tests thus avoid both
problems of predefined samples (Sect.~\ref{sec-example-based}): missed edge
cases and an implementation fitted to the samples.

\section{Hypothesis-awkward}
\label{sec-hypothesis-awkward}

Hypothesis-awkward is a collection of Hypothesis strategies that generate arrays
for testing Awkward Array and the tools that use it. This paper describes
v0.20.1~\cite{hypothesis-awkward}, released on \mbox{2026-09-22}. This version
generates arrays with nearly all possible layouts, at every level of nesting.

An earlier project~\cite{choudhury2021} in 2021 used Hypothesis to generate
inputs for the kernels of Awkward Array. In contrast, Hypothesis-awkward
generates the arrays themselves.

\subsection{The main strategy \texttt{arrays()}}
\label{sec-arrays}

The main strategy of Hypothesis-awkward is \texttt{arrays()}. Its options are
keyword arguments, each with a default value. When it is called without any, it
generates nearly all possible Awkward Arrays (Listing~\ref{lst-arrays-output}).

\begin{lstlisting}[language={}, basicstyle=\ttfamily\scriptsize,
caption={Examples of arrays generated by \texttt{arrays()} without options,
  printed with \texttt{repr()}.},
label={lst-arrays-output}]
<Array ['', '\U000c2f9f', ..., '@ú\x94j\U000c4364e'] type='4 * string'>
<Array [[], [], None, [], ..., [], [], None] type='42 * option[var * ?bytes]'>
<Array [??, ??, ??, ??, ??, ??] type='6 * var * unknown'>
<Array [NaT, NaT, ..., -9223372036854773681] type='26 * datetime64[Y]'>
<Array [[??, ??], [??, ??], ..., [??, ??]] type='8 * 2 * var * timedelta64[fs]'>
<Array [[[[[], [], [], [], []]]]] type='1 * 1 * 1 * var * var * var * bool'>
<Array [[16996], [10841], ..., [10841], None] type='7 * option[1 * uint16]'>
<Array [[None]] type='1 * 1 * option[1 * option[var * int16]]'>
<Array [[]] type='1 * option[var * 0 * union[timedelta64[D], 0 * unknown]]'>
<Array [0.0, inf, 0.0, nan, 0.0] type='5 * float16'>
\end{lstlisting}

The strategy generates arrays with all twelve subclasses of \texttt{Content}
(Sect.~\ref{sec-awkward}), nested in one another: lists of variable length,
records, and unions, with missing values represented by any of the four
subclasses, offsets that need not start at zero, and lengths of zero at every
level. The leaves have any NumPy dtype that Awkward Array supports, with values
that include \texttt{nan}, infinities, and \texttt{NaT}, or they are strings or
bytestrings. The arrays can be virtual, shown as two question marks in
Listing~\ref{lst-arrays-output}. The current version cannot generate layouts
with some special parameters, such as those of categorical data.

The strategy \texttt{arrays()} has more than 20 options to control the
output~\cite{hypothesis-awkward}. By default, every kind of layout is allowed,
and only the total size is constrained, by the option \texttt{max\_size} (50 by
default). The total size counts the scalars in all layouts in the array,
including data values, offsets, indices, and field names. Other options can
constrain the size of each leaf, the depth of nesting, the length of the array,
and the number of fields of a record. The option \texttt{dtypes} takes a
strategy for the NumPy dtypes of the leaves, and \texttt{allow\_nan=False}
disables \texttt{nan} and \texttt{NaT}. Each kind of layout can be disabled
separately with its \texttt{allow\_*} option, for example,
\texttt{allow\_union=False} for unions. With these options, a test can restrict
the strategy to the arrays that the code is expected to accept.

The strategies from which \texttt{arrays()} draws internally are also public
API: \texttt{contents()} generates a layout, and there is a strategy for each
subclass of \texttt{Content} and for the NumPy dtypes and arrays that Awkward
Array supports. They can be used directly: for example, to test code that
operates on layouts or to compose a custom strategy.

\subsection{Generating and shrinking samples}
\label{sec-generating}

The strategy \texttt{arrays()} generates the layout recursively from the root
node. It draws the subclass of \texttt{Content} for each node before drawing its
children such that the node can hold them. The strategies are designed so that
a failing array shrinks (Sect.~\ref{sec-pbt}) toward a simpler layout: fewer
elements, the dtype \texttt{bool}, and a leaf at each node, which reduces the
nesting.

\subsection{Development and testing of the strategies}
\label{sec-development}

We develop Hypothesis-awkward by test-driven development with AI coding agents,
mainly Claude Code~\cite{claude-code}. We specify the behavior of a
strategy---the agent first refactors the code to prepare for the implementation,
without modifying the tests; then generates tests that fail; and finally
generates the implementation that makes them pass. This is the cycle of
Sect.~\ref{sec-reliable} with an agent making the changes, and it depends on
reliable tests.

We test the strategies in Hypothesis-awkward with three kinds of tests:
property-based tests, reachability tests, and shrinking tests. The code coverage
is at or near 100\%. It is not easy to pass these tests unless the
implementation is correct, so we are confident in releasing a change that passes
them.

The property-based tests first draw the options of the strategy under test from
another strategy, then draw a sample from the strategy under test with the drawn
options, and finally assert that the sample satisfies the specification of the
strategy for these options. Listing~\ref{lst-strategy-test} shows the test of
\texttt{numpy\_array\_contents()}, a strategy for the leaves. In the listing,
\texttt{st\_ak} is the module \texttt{hypothesis\_awkward.strategies},
\texttt{ak} is \texttt{awkward}, and the decorator \texttt{given} of Hypothesis
turns the test function into a property-based test, which it runs many times
with different samples drawn from strategies; the strategy \texttt{st.data()}
lets the test draw from other strategies step by step.

The \emph{reachability tests} assert that the strategy can generate a given edge
case, for example, an empty list. The \emph{shrinking tests} assert that
Hypothesis shrinks the samples of the strategy to the expected simplest sample.

\begin{lstlisting}[caption={The property-based test of
\texttt{numpy\_array\_contents()}, adapted from Hypothesis-awkward.},
label={lst-strategy-test}]
@given(data=st.data())
def test_properties(data: st.DataObject) -> None:
    kwargs = data.draw(numpy_array_contents_kwargs())
    result = data.draw(st_ak.contents.numpy_array_contents(**kwargs))
    assert isinstance(result, ak.contents.NumpyArray)
    min_size = kwargs.get('min_size', 0)
    max_size = kwargs.get('max_size', DEFAULT_MAX_SIZE)
    assert min_size <= len(result) <= max_size
    if not kwargs.get('allow_nan', True):
        assert not any_nan_nat_in_numpy_array(result.data)
    # ... more assertions
\end{lstlisting}

\section{Testing Awkward Array with Hypothesis-awkward}
\label{sec-testing-awkward}

Testing of Awkward Array with Hypothesis-awkward began on 2026-02-19, when the
first property-based test was added. The goal is to test every testable property
of Awkward Array, including those of the operations (functions such as
\texttt{ak.flatten()}), of slicing (\texttt{\_\_getitem\_\_()}), and of the
kernels. The tests run in the continuous integration (CI) of Awkward Array.

\subsection{The tests so far}
\label{sec-tests-so-far}

As of 2026-09-22, the tests cover a small part of Awkward Array: fewer than 20
of about 130 operations, and only some of their properties. The tests are of
three kinds.

\begin{description}
\item[Calls with drawn options.] The test draws an array and options for which
  the operation is expected to return, then calls the operation with them; the
  call must return without raising an error. This test is a single call, yet it
  tests many lines of code through many internal calls: an operation of Awkward
  Array dispatches on the layout, so the call executes a different path for each
  of the many layouts that the strategy generates. So far, this kind of test
  exists for \texttt{ak.flatten()}, \texttt{ak.ravel()}, \texttt{ak.all()}, and
  \texttt{ak.any()}.
\item[Tests of \texttt{ak.array\_equal()}.] When the expected array is known, as
  in a round trip, a test asserts with \texttt{ak.array\_equal()} that the
  returned array equals it, so \texttt{ak.array\_equal()} itself must be tested.
  So far, two tests assert that it is reflexive and symmetric: an array is equal
  to itself, and the order of the two arguments does not affect the result. They
  also assert that \texttt{ak.array\_equal()} returns without raising an error.
  The tests currently do not draw options.
\item[Round trips of function pairs.] An array that another format can
  represent, converted to that format and then back, should equal the original.
  Listing~\ref{lst-roundtrip} shows the test of \texttt{to\_numpy()} and
  \texttt{from\_numpy()}. The test restricts the strategy with the
  \texttt{allow\_*} options to the arrays that a round trip through NumPy
  returns unchanged: it allows only rectilinear arrays of a single NumPy dtype,
  without missing values. The tests apply the same property to Arrow, Parquet,
  Feather, and JSON; each restricts the strategy to the arrays that the format
  can represent, with the \texttt{allow\_*} options and with a strategy for the
  dtypes.
\end{description}

\begin{lstlisting}[caption={The round-trip test of \texttt{to\_numpy()} and
\texttt{from\_numpy()}, adapted from Awkward Array.},
label={lst-roundtrip}]
@given(a=st_ak.constructors.arrays(
    allow_list=False, allow_record=False, allow_union=False,
    allow_string=False, ...))  # and seven more options
def test_roundtrip(a: ak.Array) -> None:
    n = ak.to_numpy(a)
    returned = ak.from_numpy(n)
    assert ak.array_equal(a, returned, equal_nan=True)
\end{lstlisting}

\subsection{Running in CI}
\label{sec-ci}

On every change to Awkward Array, the tests run with 200 samples per test and a
seed that Hypothesis fixes for each test, so a change does not fail the tests on
a sample unrelated to it; for the same reason, the tests pin Hypothesis-awkward
to an exact version. Every night, the tests run again with 10,000 samples per
test and a new seed. Arrays that trigger a known bug are excluded from the tests
until the change that fixes the bug removes the exclusion.

\section{Bugs found by Hypothesis-awkward}
\label{sec-bugs}

As of 2026-09-22, the tests have found 37 bugs in Awkward Array, of which 11 are
fixed. They have also found one bug in PyArrow~\cite{arrow} and one in
Hypothesis; both are fixed. (Numbers such as \#4316 are issues or pull requests
in \texttt{github.com/scikit-hep/awkward}.)

\subsection{Bugs in Awkward Array}
\label{sec-bugs-in-awkward}

Table~\ref{tab-bugs} lists the bugs in Awkward Array by the operation involved.
Each issue contains a reproducer with the simplest failing array that was found.

\begin{table}[h]
\centering
\caption{Bugs found by the tests as of 2026-09-22, grouped by the operation that
has the bug.}
\label{tab-bugs}
\begin{tabular}{llrr}
\toprule
Operation & Examples & Found & Fixed \\
\midrule
Equality (\texttt{ak.array\_equal()}) & \#3888, \#3921, \#3962, \#4316 & 4 & 3 \\
NumPy conversion & \#4217, \#4226, \#4227 & 3 & 0 \\
Arrow conversion & \#4219, \#4222, \#4228, \#4229 & 8 & 1 \\
Parquet conversion & \#4220, \#4305 & 2 & 0 \\
JSON conversion & \#4241, \#4243, \#4244 & 4 & 0 \\
Flattening and broadcasting & \#4214, \#4247, \#4261, \#4282 & 12 & 5 \\
Reducers and sorting & \#4259, \#4264 & 2 & 0 \\
Layout internals and virtual arrays & \#4126, \#4288 & 2 & 2 \\
\midrule
Total & & 37 & 11 \\
\bottomrule
\end{tabular}
\end{table}

\begin{description}
\item[\texttt{ak.array\_equal()}.] Four bugs were found in
  \texttt{ak.array\_equal()}, and three are fixed; the fourth, a wrong result
  for some equal union arrays (\#4316), is open.
\item[Conversions.] Of the 37 bugs, 17 are in the conversions to and from NumPy,
  Arrow, Parquet, and JSON. A conversion has to represent every layout in the
  data model of another system.
\item[Edge cases.] The arrays that trigger most of the bugs are edge cases: they
  have empty unions, lists of length zero, missing values at unusual positions,
  offsets that do not start at zero, or uncommon dtypes such as
  \texttt{datetime64[D]} and \texttt{float16}.
\item[Wrong results without errors.] Of the 37 bugs, 24 appeared only as raised
  errors, before any assertion. In the other 13, the operation returned a wrong
  result, in 5 of them in addition to an error on other arrays: for example,
  \texttt{ak.to\_arrow()} corrupted \texttt{datetime64[D]} values (\#4219), and
  \texttt{ak.flatten()} with \texttt{axis=None} dropped data on nested unions
  (\#4214).
\end{description}

\subsection{Bugs outside Awkward Array}
\label{sec-bugs-outside}

\paragraph{PyArrow.} The Feather round trip found a bug in PyArrow's writer of
the Arrow IPC file format (\#4238); we reported it with a reproducer without
Awkward Array, and it has since been fixed.

\paragraph{Hypothesis.} The strategies reached an edge case in Hypothesis
itself: an internal assertion error after shrinking, which was reduced to a
reproducer without Awkward Array, reported, and fixed in Hypothesis 6.152.4
(\#3995).

\section{Summary and outlook}
\label{sec-summary}

Hypothesis-awkward is a Python package that extends Hypothesis with strategies
for Awkward Array. The main strategy generates arrays with nearly all possible
layouts. Property-based tests that draw from the strategy run in the CI of
Awkward Array. The tests cover only one or two properties of each of fewer than
20 functions of Awkward Array; however, they have already found more than 30
bugs in it and one in Hypothesis itself. We expect many more bugs to be found as
the tests grow to cover every testable property of Awkward Array.

We plan to develop strategies that generate arrays of a given type, which can be
specified, for example, by a type string such as \texttt{var * float64}. The
current version of Hypothesis-awkward generates arrays in terms of their
layouts, which suits the testing of Awkward Array. Generated arrays of a given
type would be more useful for testing scientific tools and analysis code that
use Awkward Array, because such code expects arrays of certain types.

Many public functions of Awkward Array are each an entry point to much of its
internal code, and the property-based tests can assert properties of the entry
points independently of the details of the internal code. When the
property-based tests cover every testable property of the public functions, and
the tests of Awkward Array as a whole reliably detect any change in its external
behavior, an AI coding agent could carry out a large refactoring of the internal
code, as agents already do in the much smaller codebase of Hypothesis-awkward.
More generally, when the tests capture every expected external behavior of the
software, AI agents could generate alternative implementations that work as
drop-in replacements for the original.

Hypothesis-awkward helps make the tests of Awkward Array and of the tools that
use it reliable; reliable tests enable iterative development, shortened by AI
coding agents, and frequent releases as requirements change.

\section*{Acknowledgements}
\begin{acknowledgement}%
This work was supported by the National Science Foundation under Cooperative
Agreement \mbox{PHY-2323298} (IRIS-HEP).
\end{acknowledgement}

\bibliography{references}

\begin{thebibliography}{8}

\bibitem{fowler2018}
M.~Fowler, {Refactoring: Improving the Design of Existing Code}, 2nd~edn.
  (Addison-Wesley, Boston, MA, USA, 2018)

\bibitem{awkward}
J.~Pivarski, I.~Osborne, I.~Ifrim, H.~Schreiner, A.~Hollands, A.~Biswas,
  P.~Das, S.~Roy~Choudhury, N.~Smith, M.~Goyal, P.~Fackeldey, I.~Krommydas,
  A.~Rios-Tascon, T.~Sakuma, {Awkward Array}, \doiwoc{10.5281/zenodo.4341376}
  (2018)

\bibitem{numpy}
C.R. Harris, K.J. Millman, S.J. van~der Walt, R.~Gommers, P.~Virtanen,
  D.~Cournapeau, E.~Wieser, J.~Taylor, S.~Berg, N.J. Smith, R.~Kern, M.~Picus,
  S.~Hoyer, M.H. van Kerkwijk et~al., {Array programming with NumPy}, Nature
  \textbf{585}, 357 (2020). \doiwoc{10.1038/s41586-020-2649-2}

\bibitem{MacIver2019Hypothesis}
D.R. MacIver, Z.~Hatfield-Dodds, {many other contributors}, {Hypothesis: A new
  approach to property-based testing}, Journal of Open Source Software
  \textbf{4}, 1891 (2019). \doiwoc{10.21105/joss.01891}

\bibitem{hypothesis-awkward}
T.~Sakuma, {Hypothesis-awkward v0.20.1 [software]},
  \doiwoc{10.5281/zenodo.22896927} (2026)

\bibitem{choudhury2021}
S.~Roy~Choudhury, {Automating Awkward Array testing}, PyHEP 2021,
  \url{https://indico.cern.ch/event/1019958/contributions/4418489/} (2021)

\bibitem{claude-code}
{Anthropic}, {Claude Code}, \url{https://claude.com/product/claude-code}

\bibitem{arrow}
{Apache Arrow developers}, {Apache Arrow}, \url{https://arrow.apache.org/}

\end{thebibliography}

\end{document}